\documentclass[aps, prd, twocolumn, tightenlines, notitlepage, superscriptaddress, nofootinbib, preprintnumbers, floatfix, showkeys,10pt]{revtex4-2}
\usepackage{graphicx}
\usepackage{orcidlink}
\usepackage[normalem]{ulem}
\usepackage{amssymb}
\usepackage{amsmath}
\usepackage{graphicx}
\usepackage{url}
\usepackage{ulem}
\usepackage[utf8]{inputenc}
\usepackage{slashed}
\usepackage{lipsum}
\usepackage[T1]{fontenc}
\usepackage{appendix}
\usepackage{csquotes}

\pdfoutput = 1

\begin{document}

\title{Dark photon searches in the photon channel}
\author{D. Aristizabal Sierra\orcidlink{0000-0001-5429-3708}}%
\email{daristizabal@uliege.be}%
\affiliation{Universidad T\'ecnica Federico Santa Mar\'{i}a-Departamento de F\'{i}sica\\Casilla 110-V, Avda. Espa\~na 1680, Valpara\'{i}so, Chile}%

\author{A. Betancur\orcidlink{0000-0002-3196-1774}}%
\email{amalia.betancur@eia.edu.co}%
\affiliation{Grupo F\'isica Te\'orica y Aplicada, Universidad EIA,
A.A. 7516, Envigado, Colombia}%

\author{K. Pohl\orcidlink{0009-0009-1228-3339}}%
\email{kpohl@usm.cl}%
\affiliation{Universidad T\'ecnica Federico Santa Mar\'{i}a-Departamento de F\'{i}sica\\Casilla 110-V, Avda. Espa\~na 1680, Valpara\'{i}so, Chile}%

\author{J. Velez\orcidlink{}}%
\email{julivel08@gmail.com}%
\affiliation{Department of Physics and Astronomy, Purdue University, 525 Northwestern Avenue, West Lafayette, IN 47907, USA}%
\affiliation{Grupo F\'isica Te\'orica y Aplicada, Universidad EIA, A.A. 7516, Envigado, Colombia}%

\begin{abstract}
Spectral shape differences between photons produced in $\pi^0\to\gamma+\gamma$ and $\pi^0\to\gamma+A_D$
may provide a new avenue for dark photon searches. Assuming 70 $\mu$m thick tungsten foils separated by 200 $\mu$m and a 1 GeV proton beam, we developed a GEANT4 model to estimate photon
production and detection including background. Our results demonstrate that multiple campaign runs with a 10-50 $\mu$A beam could probe previously unexplored regions of parameter space in models where the dark photon has predominantly invisible decays. The results are highly model independent.
\end{abstract}
\maketitle

\textbf{Introduction.} Beam-dump experiments provide the conditions to test dark sector models, in particular light dark matter (LDM) scenarios \cite{Batell:2009di}. A high-intensity electron or proton beam impinged on a highly dense target material can produce directly, or indirectly through meson decays or Bremmstrahlung processes, dark sector particles. In vector portal models at GeV energies, the dark photon can decay either visibly or invisibly. Signals are thus Standard Model (SM) fermion or DM pairs. Visible modes in the form of resonances are searched with tracking systems and e.g. electromagnetic calorimeters. Dedicated experiments such as DarkQuest, NA64, SHiP, FASER, HIKE, CODEX-b, and HPS have the capability for such searches \cite{Batell:2020vqn,Gninenko:2013rka,Andreas:2013lya,Alekhin:2015byh,Feng:2008ya,HIKE:2022qra,CODEX-b:2019jve,Baltzell:2022rpd}. Because of the signals they predict, these scenarios are tightly constrained.

Models where the dark photon decays predominantly to DM pairs, instead, can be tested with the same technology employed in neutrino detection. From that perspective, dedicated accelerator neutrinos facilities provide a suitable environment to test these types of scenarios (see e.g. \cite{deNiverville:2016rqh}). Dark sector particles can be potentially produced in the collision of a proton beam on a fixed target. In vector portal models, dark photon decays to DM pairs produce a boosted DM flux that can be further detected with the same detectors used for neutrino physics. Depending on the collision energy, different production mechanisms operate: Meson decays, proton Bremsstrahlung and Drell-Yang processes. Neutrino experiments with capabilities for such searches include COHERENT, CCM, NOVA, DUNE and DUNE-PRISM \cite{COHERENT:2015mry,CCM:2021leg,NOvA:2025ykl,DUNE:2021tad,DUNE:2020fgq}. Deployment of dedicated neutrino detectors at ESS, J-PARC and CSNS will offer as well opportunities for this type of searches \cite{AristizabalSierra:2026jgp}.

Sensitivities depend on the kinetic mixing strength ($\epsilon$), the dark photon and DM masses ($m_{A_D}$ and $m_\phi$), and the dark sector fine-structure constant. Constraints from experimental data are usually presented in terms of an \enquote{effective} coupling and $m_\phi$, specializing results to particular benchmark models. From that perspective, limits are model-dependent. Following phenomenological analysis \cite{Dutta:2019nbn}, dark sector searches in the MeV regime have been conducted by the COHERENT and the CCM collaborations \cite{COHERENT:2022pli,COHERENT:2021pvd,CCM:2021yzc,CCM:2021leg}. Further analyses using $\nu-^{12}\text{C}$ inelastic scattering data have proven to provide relevant limits as well \cite{Dutta:2023fij}. Although at a different energy scale, SBND and DUNE forecasted sensitivities from $\pi^0$ decays have been calculated in Ref. \cite{Dutta:2026dvg,DeRomeri:2019kic}.

To the best of our knowledge, in the region $m_{A_D}\lesssim m_{\pi^0}$ and for a dark photon decaying invisibly, the most competitive limits in the $m_{A_D}-\epsilon$ parameter space have been placed by NA64 at the CERN SPS \cite{NA64:2016oww,NA64:2023wbi}. Results in Ref. \cite{NA64:2016oww} followed from dark photon Bremsstrahlung emission with $2.75\times 10^9$ electrons-on-target (EoT). The search relied on missing energy signals stemming from dark photon decays to DM pairs. Results in Ref. \cite{NA64:2023wbi} adopted the same strategy, but combining the statistics from the 2016-2022 runs with overall $9.37\times 10^{11}$ EoT. In the region of interest, limits from BaBar \cite{BaBar:2017tiz} have been surpassed by the latest results from NA64. In contrast to limits derived from LDM signals, these limits depend solely on $\epsilon$ and $m_{A_D}$.

In this letter, we point out that searches for dark photon signals may be done as well by looking into the photon spectra produced by $\pi^0$ decays. This requires a proper isolation of the photons produced in these decays, a clean detection technique, and a good understanding of possible backgrounds. Inspired by measurements of the $\pi^0$ lifetime carried out at CERN SPS \cite{Atherton:1985av}, we suggest that those requirements are met by an arrangement consisting of a 10-50 kW proton beam striking a $70\,\mu\text{m}$ thick tungsten foil at $\sim 1\,\text{GeV}$. Detection of the photons produced in $\pi^0$ decays is done with the aid of a second identical foil located at $200\,\mu\text{m}$ from the first one. Positrons produced in photon conversion in the field of nuclei enable doing so.

We developed a \texttt{GEANT4} model to determine the feasibility of this arrangement. In particular, the photon and positron spectra produced by interactions in the first foil and the resulting positron spectra in the second foil. These results enabled the identification of relevant positron background. The model has been validated with COHERENT simulated data on $\pi^0$ production \cite{COHERENT:2021yvp}. By rescaling these results to two benchmark values for the protons-on-target (POT), $10^{21}$ and $10^{22}$, we provide a best-case estimate of sensitivities. We show that this approach allows testing regions not yet entirely explored.

\textbf{Theoretical framework and photon spectra.}
Phenomenological aspects of dark photon scenarios are given by the following Lagrangian terms \cite{Boehm:2002yz,Pospelov:2007mp,Feng:2008ya}
\begin{equation}
    \label{eq:Lag_mass_eigenstate}
    - \mathcal{L}\supset -\epsilon\,e\,j^\text{EM}_\mu A_D^\mu + g_Dj_\mu^X A_D^\mu
    + \tan\theta_W\,\epsilon\,g_D\,j_\mu^X\,Z^\mu\ .    
\end{equation}
Here $\epsilon\equiv \epsilon^\prime\cos\theta_W$, $\theta_W$ refers to the weak mixing angle, $\epsilon$ to the kinetic mixing strength coupling, $g_D$ to the dark sector coupling and $j^\text{EM}_\mu$ and $j_\mu^X$
to the EM and dark sector currents.

Conventional searches are based on visible and invisible channel signals. In the former case decay modes of the dark photon to SM fermion pairs, while in the latter traces left by the DM through its interaction with electrons or protons. At sufficiently low collision energies, $\mathcal{O}(T_p)\simeq 1\,$GeV,  the main production mechanism is neutral pion decays. This production mechanism is particularly relevant in spallation neutron sources where hadron activity is limited to unflavored mesons. In the regime $m_{A_D}< m_\pi^0$ with $\Gamma_{A_D}^\text{Tot}\ll m_{A_D}$ (where the narrow-width approximation is valid), $\pi^0$ meson decays produce an on-shell dark photon that further decays to a DM pair. Searches in low-threshold coherent elastic neutrino-nucleus scattering (CE$\nu$NS) detectors, rely on this discovery mode.

In addition to $A_D^\mu$, the $\pi^0$ decay mode involves as well a final-state photon (\textit{semi-invisible mode}). In the $\pi^0$ decay rest frame the photon energy spectrum per decay reads
\begin{equation}
    \label{eq:partial_decay_width_pi0_RF}
    \frac{d\Gamma(\pi^0\to \gamma + A_D)}{dE^*_\gamma}=
    \Gamma(\pi^0\to\gamma+A_D)\;\delta(E_\gamma^*-E_\gamma^0)
    \ ,
\end{equation}
where $E_\gamma^*$ refers to the final-state photon in that frame and the monochromatic line is located at  $E_\gamma^0=(m_{\pi^0}^2 - m_{A_D}^2)/2/m_{\pi^0}$. The total decay width follows from the full visible mode decay width $\Gamma(\pi^0\to \gamma+\gamma)$, enlarging phase space by a factor of 2 and accounting for the massive final-state dark photon, namely
\begin{equation}
    \label{eq:Gamma_g+AD}
    \Gamma(\pi^0\to\gamma+A_D)=2\epsilon^2
    \Gamma(\pi^0\to\gamma+\gamma)\left(1-\frac{m_{A_D}^2}{m_{\pi^0}^2}\right)^3\ .
\end{equation}
From Eq. \eqref{eq:partial_decay_width_pi0_RF}, one can see that the photon spectrum in the semi-invisible mode consist of a monochromatic line shifted towards small energy values as the dark photon mass increases. In that frame one could---in principle---distinguish the photon emitted in the full visible mode from the one emitted in the semi-invisible mode, in particular when the line drifts significantly away from $m_{\pi^0}/2$.

In the laboratory frame, because of the boosting effect, the photon spectrum per decay shifts according to
\begin{equation}
    \label{eq:Gamma_Lab}
    \frac{d\Gamma(\pi^0\to\gamma+A_D)}{dE_\gamma^\text{Lab}}=\frac{\Gamma(\pi^0\to\gamma+A_D)}{2}
    \Delta E_\gamma\ .
\end{equation}
Here $\Delta E_\gamma=E_\gamma^\text{Lab}|_\text{max} - E_\gamma^\text{Lab}|_\text{min}=2E_\gamma^0\gamma_{\pi^0}\beta_{\pi^0}$, with $E_\gamma^\text{Lab}|_\text{max,min} =E_\gamma^0\gamma_{\pi^0}(1\pm \beta_{\pi^0}\cos\theta)$ ($\theta$ being the angle between the photon in the $\pi^0$ rest frame and the $\pi^0$ propagation direction). The boost variables are given by $\gamma_{\pi^0}=E_{\pi^0}/m_{\pi^0}$ and $\beta_{\pi^0}^2=(\gamma_{\pi^0}^2-1)/\gamma_{\pi^0}^2$. Thus, in that frame the line is shifted towards larger energy values by about a factor $\gamma_{\pi^0}$. Per decay, one could as well distinguish the photon line from the full visible mode from that from the invisible mode.

In a realistic case, where a high luminosity proton beam collides against a fixed dense target, a large number of $\pi^0$ mesons are produced. In the laboratory frame, their azimuthal distribution is isotropic because of spin conservation. The polar distribution will instead tend to be anisotropic with most of them clustering around the proton beam axis, depending on collision energy. In all cases the photon spectrum will broaden within $\Delta E_\gamma$ and centered around $E_\gamma^0$. The heavier the dark photon, the narrower the broadening. The spectra of the photon produced in the semi-invisible mode thus differs from the spectra of the photons produced in the full visible mode. The peak shifts towards low energies.

\begin{figure}[h]
    \centering
    \includegraphics[scale=0.55]{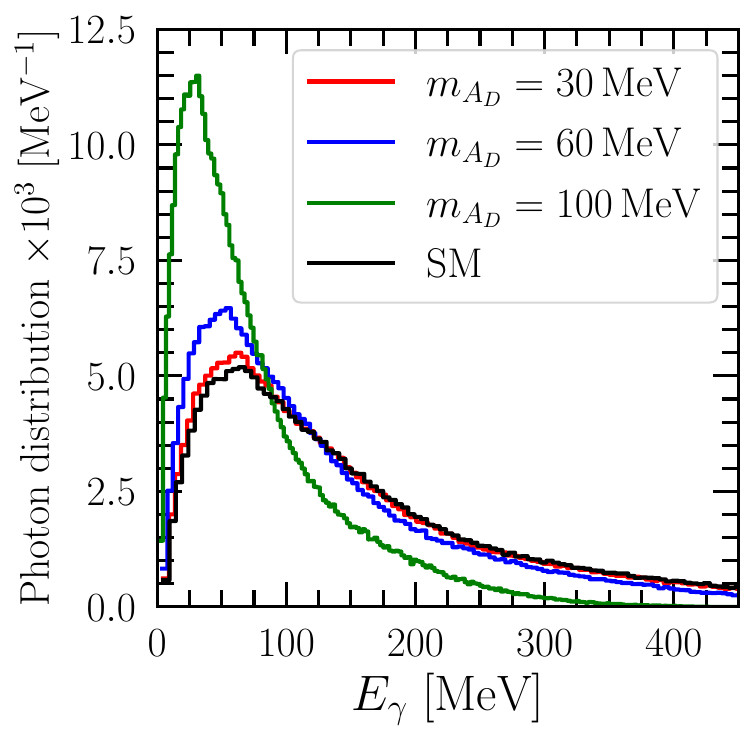}
    \caption{Photon energy distribution at production before branching fractions are accounted for.}
    \label{fig:photon_spectra_at_production}
\end{figure}

With the aim of illustrating this behavior, we have generated a $5\times 10^5$ $\pi^0$ meson sample, using the Burman \& Smith parametrization of the $\pi^+$ double differential cross section in kinetic energy and solid angle \cite{Burman:1989ds}. We have assumed a tungsten target with $T_p=1\,\text{GeV}$. From the $\pi^0$ four-momenta sample we then calculate the photons energy distributions for the SM decay mode as well as for several dark photon masses. The result is displayed in Fig. \ref{fig:photon_spectra_at_production}. Note that results from this parametrization follow closely those obtained in full simulations \cite{COHERENT:2021pvd}. Furthermore, our main results are derived using a full simulation treatment.

Relevant for this discussion is the photon \enquote{contamination}. In a proton fixed-target collision there are multiple processes that can produce photons. To limit the possibilities one could focus on collisions at low energies, e.g. 1 GeV proton kinetic energies. This significantly reduces the possible mechanisms, yet not only $\pi^0$ meson decays will be entirely responsible for photon production.

\textbf{Electromagnetic activity in proton fixed-target collisions.} In general, photons are produced in hadronic and EM processes. In this energy regime partonic direct photon production is negligible. Thus, hadronically produced photons follow from proton and neutron inelastic processes. There are multiple mechanisms operating in this case. Protons interacting with the target material nuclei produce nuclear excitation and/or fragmentation. 
MeV photons are emitted in de-excitation processes. Intranuclear cascades and coherent proton-nucleus Bremsstrahlung further produce radiation. Proton inelastic scattering is responsible for charged and neutral pion production as well, at this energy dominantly through $\Delta(1232)$ decays: $p+N\to N + \Delta(1232)\to N+N+\pi^0 (\pi^\pm)$. Neutrons are abundantly produced through the direct interaction of protons with the fixed target or through spallation. Those neutrons then undergo reactions with nuclei, which, through de-excitation, produce photons. Protons can scatter off atoms and induce electron knockout (delta-rays). While propagating in the target material, these high-energy electrons lose energy through Bremsstrahlung radiation.

\begin{figure}
    \centering
    \includegraphics[scale=0.55]{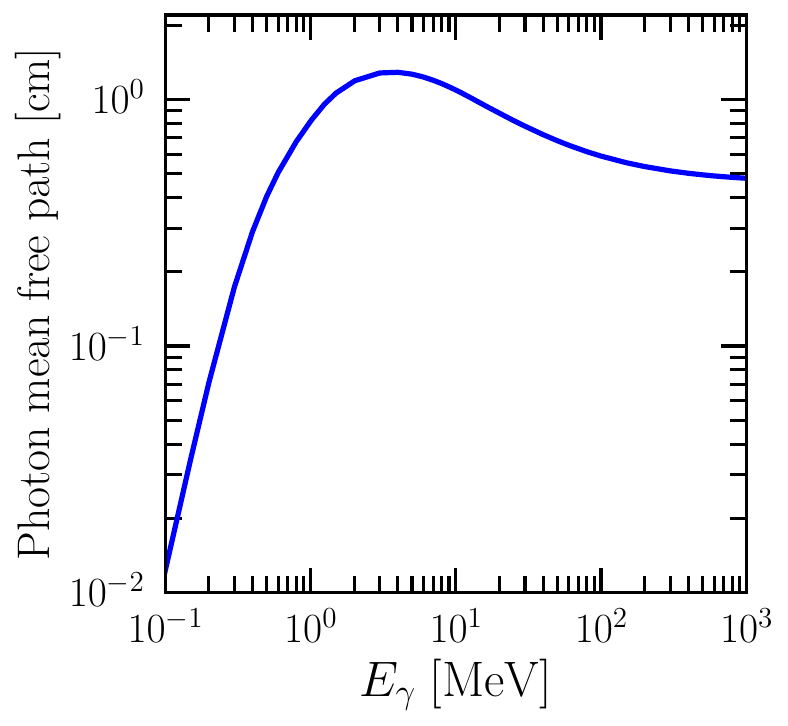}
    \caption{Photon mean free path in tungsten \cite{nist-xcom}.}
    \label{fig:photon_mean_free_path}
\end{figure}
Because of the different photo-production mechanisms, with rather sizable cross sections, the amount of photons produced is large across the whole energy range, $E_\gamma\supset [1,600]\,\text{MeV}$. The amount of radiation, however, emerging from the target depends on its type and size. Statistics depends crucially on the material density; thus, ideally, heavy materials are required. Here, we focus on tungsten, extensively used in fixed-target facilities operating in this energy regime (typically neutron spallation facilities). As shown in Fig. \ref{fig:photon_mean_free_path}, in tungsten the photon mean free path barely exceeds $\lambda_\gamma=1\,\text{cm}$ \cite{nist-xcom}. A substantial amount of photon emission then requires thin target foils \footnote{Thin foils offer other advantages: Reduced multiple scattering and small EM shower radiation.}. The thinner the target, the larger the amount of photons that emerge. Furthermore, in thinner targets possible \enquote{contamination} processes are suppressed.

Detection of the photon flux can be done with the aid of a second foil. In tungsten, above $E_\gamma=20\,\text{MeV}$, the photon total cross section is dominated by pair production in the nuclear field. At those energies, the Compton scattering cross section is already suppressed by about a factor of 40, while the cross section for pair production in the electron field by almost two orders of magnitude \cite{nist-xcom}.
A fraction of the photons ejected from the first foil will hit the second, with the fraction determined by the foils separation. For those above 20 MeV, a fraction will undergo pair production. The conversion probability in a thin foil is small, $P=1-e^{-t/\lambda^\text{pair}_\gamma}$ ($t$ refers to the foil thickness). However, in a high-statistics environment---as those provided by neutron spallation facilities---a fair amount of positrons can be measured.

\textbf{Proof of concept and expected sensitivities.} To demonstrate the feasibility of this experimental configuration we have run a two-stage \texttt{GEANT4} simulation \cite{Allison:2016lfl} consisting of two $t=70\,\mu\text{m}$ (area 1 $\text{cm}^2$) thick foils with a separation of $200\,\mu\text{m}$. In the first stage, $7\times 10^8$  protons at $T_p=1\,\text{GeV}$ strike the first foil along the $z$ axis. The simulation has been validated against the $\pi^0$ energy and angular distributions reported by the COHERENT collaboration \cite{COHERENT:2021yvp}. For the beam and target conditions of that reference, 0.1 $\pi^0$/POT are produced. In our case, instead, the yield is 0.007 $\pi^0$/POT.

The four-momenta of the neutral pions produced in the collision are recorded and further used to construct the spectra of photons produced in $\pi^0\to\gamma+\gamma$ and $\pi^0\to \gamma+A_D$ (see Fig. \ref{fig:photon_spectra_at_production} for a few spectra samples). To handle the meson decay to $\gamma+A_D$, we used \texttt{MadGraph} \cite{Alwall:2014hca} and its plugin \texttt{MadDump} \cite{Buonocore:2019rse}. We assume that 100\% of the photons produced in these two processes exit the foil, an assumption well justified by noting that the survival probability for $E_\gamma>1\,\text{MeV}$ exceeds 99\%.

\begin{figure}[h]
    \centering
    \includegraphics[scale=0.55]{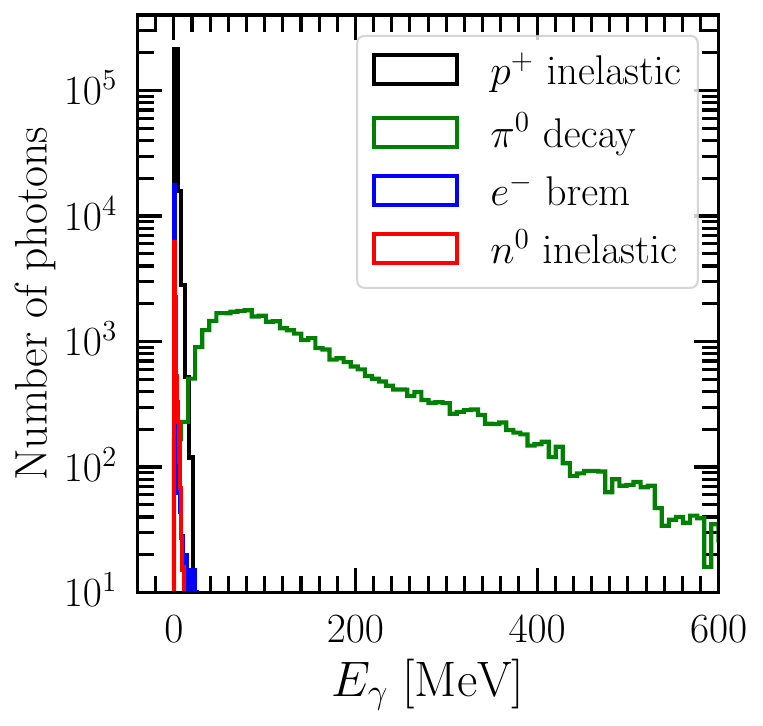}
    \caption{Photons generated in a \texttt{GEANT4} simulation with $5 \times10^6$ POT impinged on a $70\,\mu\text{m}$ thick
    tungsten foil.}
    \label{fig:photon_production}
\end{figure}

To see whether a photon flux consisting mainly of $\pi^0$ decays can be filtered, we have recorded a $\sim 3\times 10^5$ photon sample of all the exiting photons. Fig. \ref{fig:photon_production} shows the result for processes classified according to the previous discussion: proton and neutron inelastic, electron Bremsstrahlung, and $\pi^0$ decays. This result demonstrates that photons from $\pi^0$ decays can be efficiently selected by imposing the kinematic cut $E_\gamma\gtrsim 20\,\text{MeV}$.

\begin{figure}[h]
    \centering
    \includegraphics[scale=0.55]{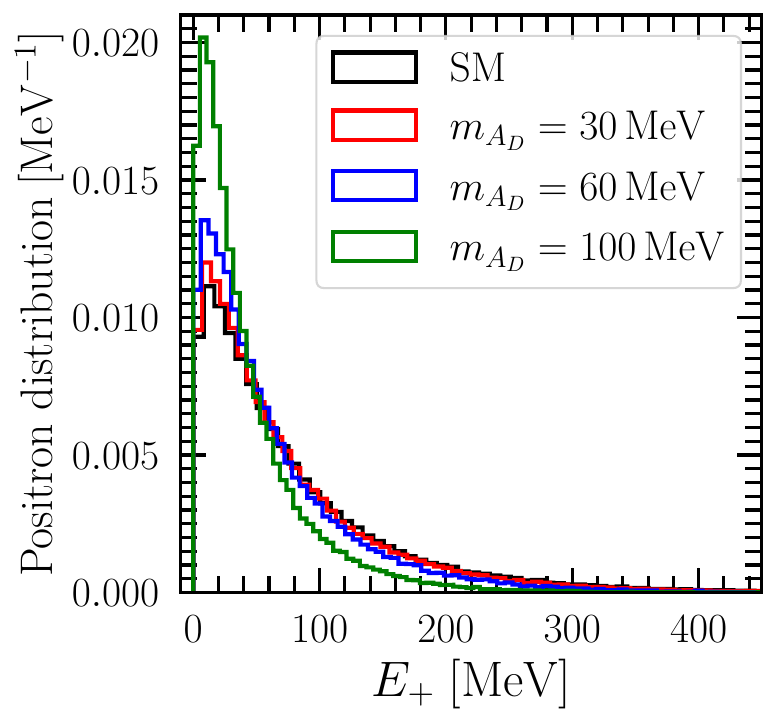}
    \caption{Positrons spectra (before weighting by branching fractions) samples obtained by photons from $\pi^0\to \gamma+\gamma$ and $\pi^0\to \gamma+A_D$ undergoing pair production in the field of atomic nuclei.}
    \label{fig:positron_spectra}
\end{figure}

In the second stage of the simulation, the spectra of the photons striking the second foil are selected through geometrical cuts along the $x$ and $y$ axes. Because of the $200\,\mu\text{m}$ separation, we find that about $70\%$ of all the photons survive the cut. The exiting positrons are finally recorded. The interaction probability through electron pair conversion in the field of atomic nuclei is of the order of $1\%$. This, however, is compensated by the high statistics implied by a large POT. Fig. \ref{fig:positron_spectra} shows a few spectra samples, including the SM spectrum, before weighting by the branching ratios.

The positrons collected in this way are expected to originate predominantly from photons produced in $\pi^0$ decays. The dominant background for signal identification arises from the standard decay channel $\pi^0 \to \gamma+\gamma$. In addition, we found that photons from $\pi^0 \to \gamma+\gamma$ and $\pi^0\to \gamma+A_D$ decays can be produced and converted in the first foil, thereby contributing to the observed positron sample. Among the other $\pi^0$ decay modes, Dalitz decay provides the next most important contribution with a branching ratio of order $1.2\times 10^{-2}$, whereas the $e^+e^-$ and $4\gamma$ modes have branching ratios below $10^{-7}$ \cite{ParticleDataGroup:2024cfk}. Therefore, the main background arises from the two $\pi^0$ most relevant decays.

There are other potential background sources generated from collisions on the first foil. A large population of protons and neutrons as well as electrons and positrons exit and strike the second foil. We have studied their spectra and found that protons generate a spectrum as that shown in Fig. \ref{fig:photon_production}. Neutrons and electrons generate mainly low energy photons, with $E_\gamma\lesssim 20\,\text{MeV}$. Most of them are generated in nuclear de-excitation or Bremsstahlung processes. 
Thus, particularly relevant for the positron background are, again, the ones produced by: the $\pi^0$ decay to $\gamma+\gamma$ with a subsequent conversion to $e^+e^-$ pairs and the $\pi^0$ Dalitz decay. With small corrections, the positron mean free path ($\lambda_-$) amounts to that of electrons. For $E_-\gtrsim 10\,\text{MeV}$ one finds $\lambda_-\gtrsim 3\,\text{mm}$ \cite{nist-pstar}. Thus, positrons with those energies have a 97\% (or higher) likelihood of contributing to background. 

Positron energy spectra are the main handle for signal identification. For that aim a compact magnetic spectrometer downstream from production could be employed for charge separation and momentum reconstruction, followed by an electromagnetic calorimeter for energy measurement. Similar setups have been used at JLAB and have been dicussed by the LUXE experiment \cite{Abramowicz:2021zja,Adrian:2022nkt}. Thus, we argue that the same strategy could be employed in the reconstruction of the positron spectra.

Given the thickness of the foils, the question of what kind of POT they can withstand for a certain data-taking time becomes relevant. The number of protons per time is determined by the beam current $I$, $\dot N=6.2\times 10^{12}\,(I/\mu\text{A})\,p^+/\text{s}$. Thus, since the collision stopping power for a 1 GeV proton in tungsten is about $23.6\,\text{MeV/cm}$ \cite{nist-pstar}, the dissipated power in the foil is $P\simeq 0.17\,\text{W}\,(I/\mu\text{A})$. This means that temperature variations over time are of the order of $dT/dt\simeq 9.3\,(I/\mu A)\,\text{K/s}$. This estimate indicates that the proton beam current must be limited to the tens of microampere range to prevent foil melting. There are as well issues related with radiation damage of the foils, but they are beyond the scope of this letter. Note that systems of rotating and cooled foils have been discussed for SHiP in Ref. \cite{ANTALIC2004185,Bharadwaj:2005hz,Liu:2016qtq}. Such setups could be in principle envisaged for this type of measurement. Based on the experience of pion lifetime measurements and other dark photon search experiments \cite{APEX:2011dww,Atherton:1985av}, we argue that the foils may be safely operated up to $\sim 10^{20}-10^{21}\,\text{POT}$. Thus, ISIS and LANSCE appear to be well-suited facilities for this experimental setup \cite{ISIS,LANSCE}.

\begin{figure}
    \centering
    \includegraphics[scale=0.55]{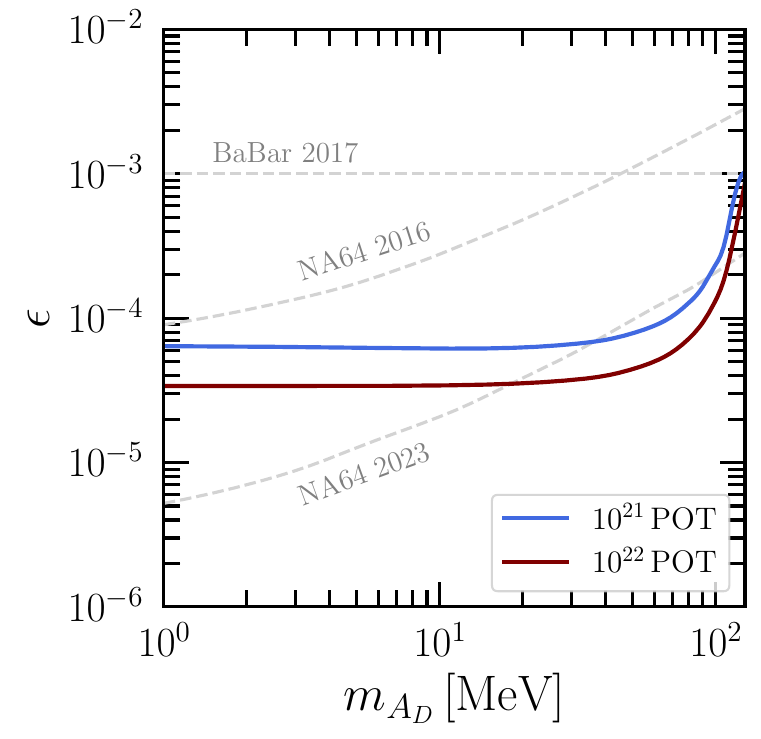}
    \caption{90\% CL limits on the kinetic mixing parameter and the dark photon mass. Limits from other experiments include NA64 2016 \cite{NA64:2016oww}, NA64 2023 \cite{NA64:2023wbi}, and BaBar 2017 \cite{BaBar:2017tiz}. Note that NA64 results are less stringent in $U_{L_\mu-L_\tau}$ models. In contrast, the technique we are pointing out does not depend on couplings to leptons. Thus, it could potentially target regions not yet explored in these type of scenarios.}
    \label{fig:sensitivities}
\end{figure}

We compare the positron spectra generated by the photons from the semi-invisible mode with the positron spectrum induced by the photons from the $\pi^0\to\gamma+\gamma$ mode. To determine best-case sensivities we rely on the following spectral $\chi^2$ function
\begin{equation}
    \label{eq:chisq}
    \chi^2=\sum_i\left(\frac{N_i^\text{obs} -N_i^\text{pred}}{\sigma_i}\right)^2\ .
\end{equation}
Here $i$ runs over positron energy bins, $N_i^\text{obs}=N_i^{\pi^0}$, $N_i^\text{pred}=N_i^{\pi^0}-\text{Br}(\pi^0 \rightarrow \gamma + A_D) \times (N_i^{\pi^0}+ N_i^\text{BSM})$, $N_i^\text{BSM}$ are the number of BSM events before weighting by the branching fraction, and $\sigma_i^2=N_i^\text{obs}$. Results are shown in Fig. \ref{fig:sensitivities}. They demonstrate that this experimental setup can probe regions of parameter space not yet explored, particularly in the range $20 \ \text{MeV} \lesssim m_{A_{D}} \lesssim 100 \ \text{MeV}$. Larger regions could be tested in scenarios where tree level couplings of the dark photon to electrons are forbidden.

\textbf{Conclusions.} By considering dark photon production in $\pi^0$ decays, we have studied the possibility of testing vector portal models through the measurement of photon spectra. Differences between the dominant di-photon decay mode and the dark sector mono-photon mode might enable doing so. We have argued that such measurement requires a relatively pure photon flux, arising mainly from $\pi^0$ decays, along with high statistics and controlled background. A low energy and moderate intensity proton beam along with an array of two thin tungsten foils, one used as target and the other to enable photon detection, met those requirements. Using a \texttt{GEANT4} model, we have demonstrated that such measurement might be possible.

Using results from the simulation, scaled to $10^{21}$ and $10^{22}$ POT, we have provided best-case estimate sensitivities. Our results show that this approach has the potential to cover unexplored regions in parameter space. This becomes particularly relevant in scenarios where the dark photon does not couple to electrons [e.g. in $U(1)_{L_\mu-L_\tau}$ models]. In this case NA64 constraints searches will be weakened, opening regions yet to be explored.

We point out that dedicated measurements of the photon activity in $\pi^0$ decays might contribute to the searches of vector portal models. We emphasize that this analysis represents a first step toward developing new strategies that rely on photon spectral measurements to uncover dark sectors.

\textbf{Acknowledgments.} We thank Paola Arias and Doojin Kim for very useful comments on this paper. The work of D.A.S. and K.P. is supported by ANID grants \enquote{Fondecyt Regular} 1221445 and 1260595. D.A.S. thanks EIA for the hospitality during the initial stage of this work. K.P. thanks as well \enquote{Programa de Iniciaci\'on a la Investigaci\'on Cient\'{i}fica (PIIC)} number 008/2025 for support. A.B. thanks Universidad T\'ecnica Federico Santa Mar\'{i}a for the hospitality during the initial stage of this work.

\bibliography{biblio}
\end{document}